\documentclass[a4paper]{jpconf}
\usepackage{graphicx}
\begin{document}
\title{A Compact TPC for the sPHENIX Experiment}

\author{Henry T. Klest on behalf of the sPHENIX Collaboration}

\address{Department of Physics and Astronomy, Stony Brook University, Stony Brook, New York 11794, USA}

\ead{henry.klest@stonybrook.edu}

\begin{abstract}
The sPHENIX detector to be installed at RHIC in 2022 is designed to precisely measure jets, jet correlations, and dilepton pairs in heavy-ion collisions. With these measurements in mind, sPHENIX will employ a compact TPC covering 20cm $< r <$ 78 cm and $|\eta| <$ 1.1 as the central tracker. Utilizing an optimized Ne-CF$_{4}$ gas mixture, zigzag readout pads, a 1.4~T solenoid, and a modified SAMPA chip for streaming readout, the TPC will provide a position resolution sufficient for measuring target observables in a high event rate environment. The design of the TPC will be discussed, as well as test beam data and potential applicability to the EIC.
\end{abstract}

\section{Introduction}

sPHENIX \cite{sPHENIX:2015aa} is a state-of-the-art detector currently under construction at the Relativistic Heavy-Ion Collider (RHIC) at Brookhaven National Laboratory. sPHENIX is designed to make measurements of jet substructure, fragmentation functions, heavy flavor, and individual $\Upsilon$(nS) states in proton and heavy ion collisions to advance understanding of the quark-gluon plasma \cite{Reed_2017}. 

Employing a MAPS-based silicon vertex detector (MVTX), an intermediate silicon tracker (INTT), and a compact TPC \cite{Klest_2020} as the tracking system, along with precision electromagnetic and hadronic calorimetry \cite{Aidala_2018}, sPHENIX will round out the RHIC physics program with high statistics jet and heavy flavor measurements \cite{Roland:2019OJ} \cite{Ji:2021tki}. 

Key to the success of sPHENIX is a combined tracking momentum resolution good enough to resolve the $\Upsilon$ states from each other in the di-electron decay channel in both $p+p$ and $Au+Au$ collisions at $\sqrt{s_{NN}}$ $=$ $200$ $GeV/c$. Along with jet substructure measurements, this sets the requirement for the sPHENIX tracking system to have a momentum resolution of $\frac{\Delta P}{P}$ $\approx$ $.02 \cdot P$ for charged particles with momenta around 5 $GeV/c$. 

In addition to the requirements on detector resolution, the sPHENIX experiment is required to take large amounts of data at a $Au+Au$ event rate of 15 kHz or greater to maximize the statistics for rare probes. This rate requirement demands that the TPC operate in an ungated mode with a fast drift velocity gas to reduce occupancy.

\section{sPHENIX TPC Design Considerations}

\subsection{Design Overview}
The sPHENIX TPC covers 20 cm $< r <$ 78 cm but is only instrumented for physics starting at 30 cm. The maximum drift length for an electron is 1.05 m in z. The readout on each endcap is segmented 36 times.  The field cage and gas volume of the TPC are merged into one structure. The field cage is constructed of five long, flexible PCBs with very fine stripes designed to step the voltage down smoothly from the central cathode to the top of the gas electron multipliers (GEMs) and produce a highly uniform electric field in the z-direction. The central cathode itself is constructed of rigid ENIG-coated FR4 segments supported by a central honeycomb structure. The cathode will be held at 45 kV and will have aluminum stripes evaporated on it for laser calibration.

\begin{figure}[ht]
\begin{minipage}{21.7pc}
\includegraphics[width=21.7pc]{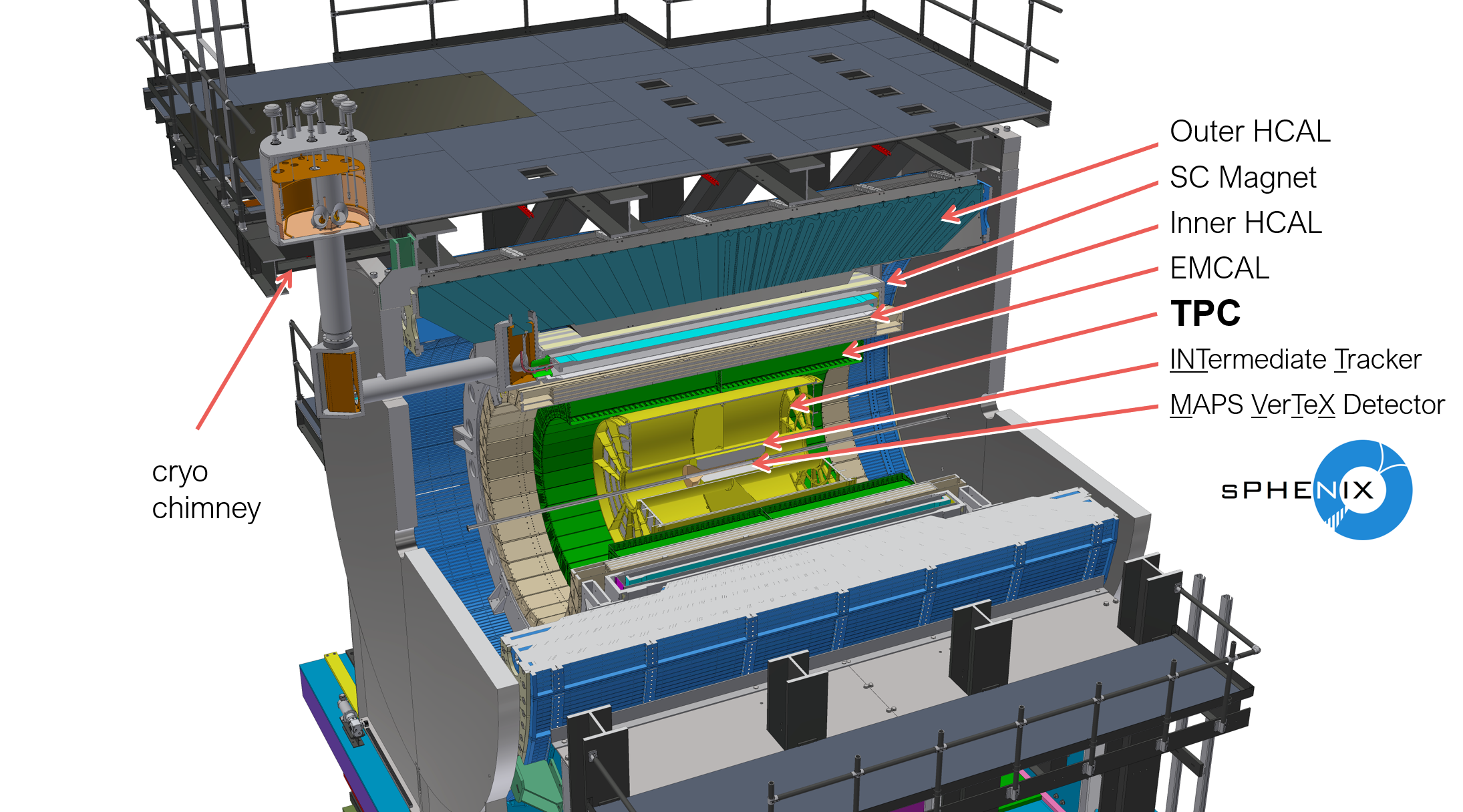}
\caption{\label{Detector}sPHENIX detector and subsystems.}
\end{minipage}
\begin{minipage}{16pc}
\includegraphics[width=16pc]{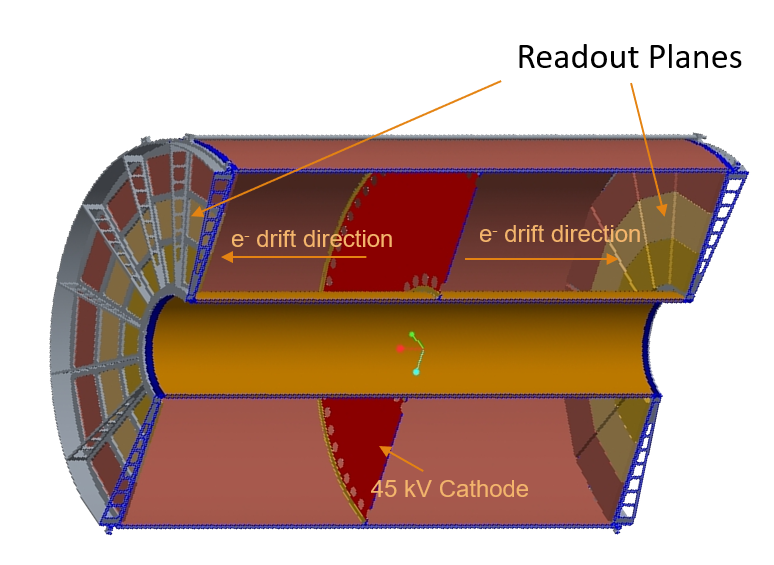}
\caption{\label{TPC}Sketch of the sPHENIX TPC.}
\end{minipage} 
\end{figure}

\subsection{Size}

The primary drivers of the size of the sPHENIX TPC are the requirements that sufficient room be allotted for calorimetry inside the magnet bore, and that the TPC sit in the uniform region of the magnetic field.

Although the radial component of the magnetic field in the TPC volume is small compared to the overall magnitude of the field, this small component causes centimeter-scale distortions to electron positions that need to be corrected for. An ionizing laser system that provides straight lines of ionization throughout the TPC will allow for straightforward calibration of this effect. 

Additionally, the electromagnetic calorimeter requires radial extension to fully contain high $p_{T}$ EM showers. The measurement and identification of electrons and photons at high $p_{T}$ enables the study of $\Upsilon$ decays and $\gamma$ + jet events, both of which are cornerstones of the sPHENIX physics program. sPHENIX is also the first midrapidity detector at RHIC to utilize a full 2$\pi$ hadronic calorimeter in the barrel. One portion of this hadronic calorimeter, the so-called inner HCal, also sits inside the magnet. This section improves the hadronic jet resolution of sPHENIX and better enables the use of particle flow techniques due to the reduced ambiguity in matching tracks to hadronic calorimeter clusters.

\subsection{Momentum Resolution}
The measurement of momentum in a TPC boils down to the ability to pinpoint the initial location of primary ionization electrons in the gas. There are many factors which affect a TPC's ability to do this, but some of the most notable are the diffusion in the gas, the intrinsic spatial resolution of the readout, the magnetic field applied, and the distortion of tracks due to space charge. The spatial resolution of a TPC can thus be parameterized as in Eq. 1.

\begin{equation} 
\sigma_{x}^{2} = \frac{D_{T}^{2}L}{N_{eff}} + \sigma_{pad}^{2} + \sigma_{Space~Charge}^{2}
\end{equation}

\subsubsection{Diffusion}
To reduce the first term of Eq. 1, the operating gas must have low diffusion. Since the sPHENIX physics program does not require particle identification by dE/dx in the TPC, the gas can be chosen to optimize position resolution. sPHENIX utilizes the 1.4~T solenoid that was designed for and operated successfully in the BaBar Experiment at SLAC. In comparison to STAR and ALICE, the magnetic field of this solenoid is quite strong. This is necessary for sPHENIX to precisely study high $p_{T}$ probes of the QGP, both due to the increased bending of tracks and the favorable effect this field has on the TPC spatial resolution. 

A strong, uniform magnetic field is especially suited for a TPC because the magnetic field decreases the diffusion of the electrons in the gas. The mechanism behind this is quite simple. As the electrons diffuse, they pick up velocity components in the directions transverse to the magnetic field. This causes them to spiral and remain pinned to the direction of the parallel $\vec{E}$ and $\vec{B}$ fields, effectively negating the transverse kicks given to the electrons as they traverse the gas. 

The diffusion will necessarily worsen the resolution for longer electron drift lengths. Thus the spatial resolution for electrons ionized near the readout will be better than those ionized near the central cathode. One of the advantages of the small size of the sPHENIX TPC is that the maximum drift length is only 1.05 m, to be compared with 2.45 m in ALICE \cite{Alme_2010} and 2 m in STAR \cite{Anderson_2003}.

The sPHENIX TPC employs a gas mixture of 50$\%$ neon with 50$\%$ CF$_{4}$ \footnote{At the time of writing of these proceedings, the gas mixture was Ne:CF4 50:50. Neon has recently become difficult to acquire, and for that reason the gas mixture was changed to Argon:CF4 in a 60:40 mixture. The properties are almost identical to those in Fig. 3, but the ion mobility is worsened.} to provide low diffusion and high drift velocity, as can be seen in Fig. 3. The longitudinal diffusion and drift velocity are independent of the magnetic field to first order. 
\begin{figure}[ht]
\begin{center}
\includegraphics[width=17.9pc]{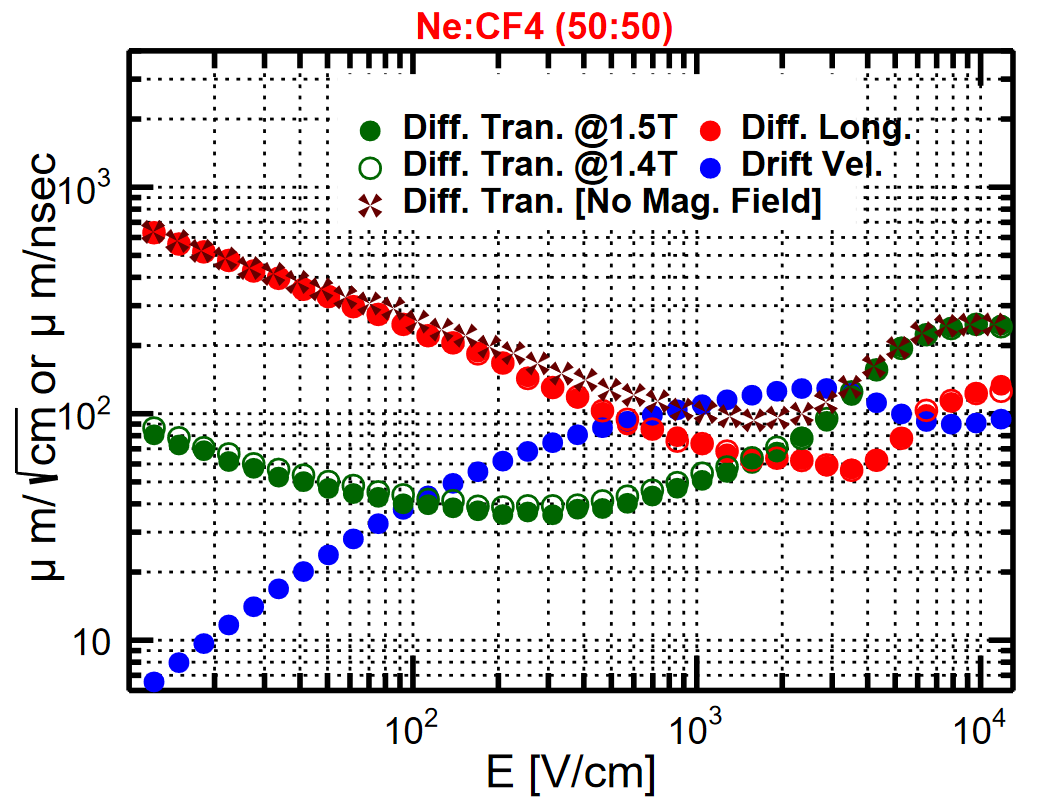}
\caption{\label{gas_description}Ne:CF$_{4}$ 50:50 gas properties. The vertical axis is a measure of diffusion or drift velocity. To minimize transverse diffusion, 400 V/cm is chosen as the operating point.}
\end{center}
\end{figure}
The drift velocity of the sPHENIX configuration is $\sim$ 80 $\mathrm{\mu m/ns}$ compared to $\sim$ 55 $\mathrm{\mu m/ns}$ for the P10 gas of the STAR TPC and $\sim$ 25 $\mathrm{\mu m/ns}$ for the Ne:CO$_{2}$:N$_{2}$ 90:10:5 gas mixture of the ALICE TPC upgrade. The high drift velocity necessitates fast electronics to have acceptable position resolution in z and low occupancy. Therefore, sPHENIX is using the 80 ns peaking time option of the SAMPA ASIC designed for the ALICE upgrade \cite{Adolfsson_2017}. The sPHENIX TPC will additionally be operated in streaming readout mode.

\subsubsection{Pad Resolution}
The second term in Eq. 1 comes from the intrinsic pad resolution. It has been shown elsewhere \cite{Azmoun_2018} that pads designed to optimally share charge can have better intrinsic position resolution than similarly sized rectangular pads. This allows for macroscopically large pads to have excellent resolution with relatively small channel counts. For this reason, the sPHENIX TPC has zig-zag patterned readout pads, shown in Fig. 5, which are designed to have charge sharing between at least two pads for all incoming charge clouds from the GEMs.

The position resolution of the sPHENIX TPC prototype was studied extensively with the 120 GeV/c proton beam at the Fermilab Test Beam Facility. The prototype is a one-sided, 40 cm drift length TPC with one full readout module installed. Due to the bunched structure of the beam and the spacing between spills, space charge contribution to the resolution was negligible. The pad and diffusion resolution terms could be extrapolated to the case where a magnetic field was present and the TPC was full length. The end result of this study showed a position resolution of 90 $\mu$m for electrons deposited just above the readout and 130 $\mu$m for electrons drifting the full 105 cm TPC drift length. 

\begin{figure}[ht]
\begin{center}
\begin{minipage}{12pc}    
\includegraphics[width=12pc]{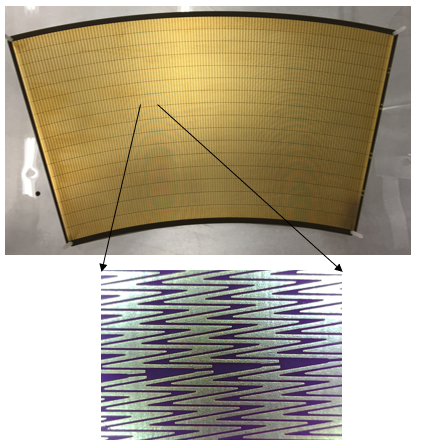}
\caption{\label{Zigzags}sPHENIX readout plane with closeup of zig-zag pads.}
\end{minipage}\hspace{2pc}%
\begin{minipage}{16pc}  
\includegraphics[width=16pc]{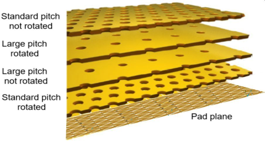}
\caption{\label{GEMs}sPHENIX Quadruple GEM layout.}
\end{minipage} 
\end{center}
\end{figure}

\subsubsection{Space Charge}
The ability of the TPC to precisely measure momentum also depends on minimization of track distortions due to electric and magnetic field non-uniformities. All TPCs have time-independent track distortions due to realistic field effects such as field cage granularity, radial $\vec{B}$ field components, etc. These distortions are typically large and remain throughout the life of an experiment, however they can typically be easily calibrated away.

A more difficult to calibrate effect is the time-dependent electric field produced by the buildup of space charge. This effect is dominated by positive ions from the gain stage that drift slowly throughout the volume of the TPC until they reach the TPC field cage or cathode and neutralize. The space charge density and track distortion near the inner field cage is very large, which informed the decision to begin physics instrumentation 10 cm beyond the inner radius of the TPC. In lieu of normal readout, this region has large pads designed to monitor the amount of incoming charge, providing additional information about the amount of ions and their locations in the TPC volume at a given time. The magnitude of the space charge distortions can be drastically reduced by decreasing the fraction of ions that escape from the GEMs without neutralizing. To reduce this ion back-flow (IBF) fraction below 1$\%$, the sPHENIX TPC uses four GEMs in an orientation similar to the ALICE upgrade \cite{Adolfsson_2021}, as shown in Fig. 6. The operating voltages of the GEMs were determined via specialized bench experiments that could measure the amount of IBF and were tested thoroughly in the sPHENIX prototype TPC in test beam.

In addition to the ionizing laser mentioned earlier, a diffuse laser calibration system shining on aluminum stripes evaporated onto the TPC central cathode will allow for real-time measurement of space charge-induced distortions as the experiment runs with beam. Aluminum was chosen because its work function is lower than that of the gold cathode, allowing photoelectrons to be liberated from the stripes but not the central cathode itself. These liberated electrons will traverse the entire drift of the TPC, and accumulate information about the distribution of space charge. By comparing the known pattern of the aluminum stripes to the detected pattern at the readout plane, one can measure the integrated distortions that electrons receive as they traverse the TPC. Care has been taken to optimize the amount of diffusion and the power output of the laser such that a uniform illumination can be achieved.

\section{Re-use at an Electron-Ion Collider Experiment}
Many of the components of sPHENIX, including the aforementioned calorimeters, can be repurposed for studying $e+p$ and $e+A$ collisions at the future Electron-Ion Collider (EIC) to be housed at the current RHIC facility. The event rate for $e+p$ collisions at the EIC will be similar to $Au+Au$ collisions in sPHENIX ($\sim$ 15 kHz), but the track multiplicity will be much lower. The average track multiplicity at the EIC will be around 15 tracks per event, compared to 400 tracks per event in $Au+Au$ events at RHIC. This means that the occupancy of the TPC will be much lower, and perhaps an optimization could be done to allow the TPC to identify hadron species via dE/dx while also improving the position resolution. Space charge will be less of an issue, so the region from 20-30 cm in r can be instrumented for physics.

\begin{figure}[ht]
\begin{center}  
  \includegraphics[scale=.6]{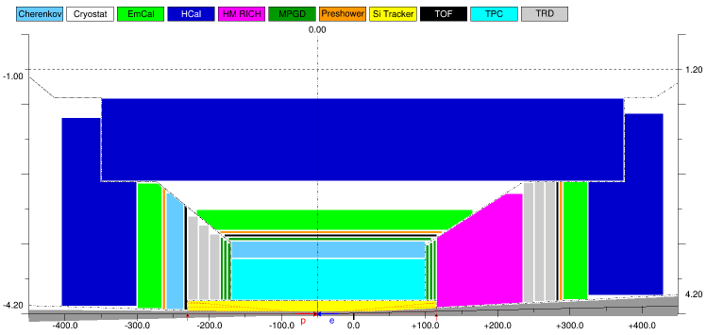}
  \caption{\label{sPHENIX}Example of a possible EIC detector configuration.}
\end{center}
\end{figure}

Another option is to modify the TPC by removing one of the endcaps and moving the cathode to the end in the electron-going direction. The impetus for this modification is that the readout of the TPC requires mechanical support and cooling, which contribute to the material budget. Since the measurement of the direction, momentum, and species of the scattered electron is of utmost importance at the EIC, it is reasonable to attempt to reduce the material in the electron-going direction. While this will increase the average drift length and worsen the TPC spatial resolution at large drift lengths due to diffusion, the reduced material will generally improve the reconstruction and identification of the scattered electron.

\section{Summary}
The compact sPHENIX TPC will be installed into sPHENIX in July 2022. The various factors affecting the TPC momentum resolution and rate capability have been discussed, along with the solutions to be employed. All studies performed, including test beams and bench experiments, indicate that the TPC will be able to achieve the resolution required by the sPHENIX physics program.

\section*{References}
\bibliographystyle{iopart-num} 
\bibliography{Refs}

\end{document}